\documentclass{llncs}
\usepackage{graphicx}
\usepackage{algorithm}
\usepackage[noend]{algorithmic}
\usepackage{subfigure}
\usepackage{amsmath} 
\usepackage{multirow}

\begin{document}
\title{Efficient Compression and Indexing of Trajectories\thanks{\scriptsize{Funded in part by European Union’s Horizon 2020  Marie Sk{\l}odowska-Curie grant agreement No 690941; MINECO  (PGE and FEDER) [TIN2016-78011-C4-1-R;TIN2013-46238-C4-3-R]; CDTI, MINECO [ITC-20161074;IDI-20141259;ITC-20151305;ITC-20151247]; Xunta de Galicia (co-founded with FEDER) [ED431G/01]; and Fondecyt Grants 1-171058 and 1-170048, Chile. }	}}

\author{Nieves R. Brisaboa\inst{1} \and Travis Gagie\inst{2} \and Adri\'an G\'omez-Brand\'on\inst{1} \and Gonzalo Navarro\inst{3} \and Jos\'e R. Param\'a\inst{1}
}
\institute{
 Universidade da Coru\~na, Computer Science Deparment, Spain. \\
\email{\{brisaboa,adrian.gbrandon,jose.parama\}@udc.es}
\and
School of Informatics and Telecommunications, Diego Portales University, Santiago, Chile.
\email{travis.gagie@mail.udp.cl}
\and
Dept. of Computer Science, University of Chile, Chile. 
\email{gnavarro@dcc.uchile.cl} }

\maketitle

\begin{abstract}

We present a new compressed representation of free trajectories of moving objects. It combines a partial-sums-based structure that retrieves in constant time the position of the object at any instant, with a hierarchical minimum-bounding-boxes representation that allows determining if the object is seen in a certain rectangular area during a time period. Combined with spatial snapshots at regular intervals, the representation is shown to outperform classical ones by orders of magnitude in space, and also to outperform previous compressed representations in time performance, when using the same amount of space.

\end{abstract}

\section{Introduction}

With the appearance of cheap devices, such as smartphones or GPS trackers, which record the position of moving objects, the  need to efficiently store and manage 
information on trajectories has become commonplace. 
Although storage, network, and processing capacities are rapidly increasing, the available data grows faster, and demands reduced-size representations \cite{zheng11}. The first option is to lose precision and discard points of the acquired trajectories, with more or less sophisticated procedures. A second choice is to keep all the points of the trajectories and use differential compression \cite{TrajStore,Trajic}. These methods store for each coordinate $(x,y)$ the difference with the previous point. The problem is that, to obtain the coordinates of the $i^{th}$ point, we must add up all the preceding differences. This is a variant of the {\em partial sums problem} where the values can be positive or negative.

Our new method, called Constant Time Access Compressed Trajectories (ContaCT), uses an Elias-Fano-based \cite{Fan71,Eli74} representation of the differences that allows computing the partial sums in constant time while using space comparable to other differential encoding methods. In addition to constant-time access to the trajectory data, ContaCT provides a hierarchical structure that allows efficiently answering time-interval queries \cite{PfoserJT00} (i.e., determine if an object is seen inside a rectangular area during a time interval) without the need to follow all the movements of the object in the queried interval. We use ContaCT to represent the trajectories of a large set of objects. At regular time instants, ContaCT includes a spatial snapshot with a structure that supports range queries, which is useful to bound the objects that must be tracked to answer time-interval queries.

Our experiments on a set of real trajectories of ships shows that, while there exist techniques based on grammar-compression that use less space than ContaCT \cite{BrisaboaGNP16}, our index is up to 2.7 times faster when using about the same amount of space. Our index is also much faster than a baseline differentially compressed representation, for about the same space. We also compared ContaCT with a classical MVR-tree, where trajectories are stored as sets of points and time-interval queries reduce to 3D range queries. It turns out that ContaCT required 1,300 times less space, and it was still faster in time-interval queries spanning more than 14 instants.

\section{Background}

A trajectory is a sequence of timestamped geographic positions in the two-dimensional space. We assume that the recorded timestamps are regularly placed over time, possibly with periods of time without values. 
We also assume that the recorded timestamps are exactly the same for all the objects.






Apart from the basic functionality of returning the whole trajectory of an object or its position at some time instant, we deal with the following, more elaborate queries \cite{PfoserJT00}: \textit{time-slice}  returns all the objects in a given query region at a given timestamp, and {\em time-interval} returns all the objects that overlap the query region at any time instant of an interval.

\medskip
\noindent
\textbf{Bitmaps.}
A \textit{bitmap} is a binary sequence $B[1,n]$ that supports the following operations: (i) $access(B,i)$ returns the bit $B[i]$, (ii) $rank_b(B,i)$ returns the number of occurrences of bit $b\in{0,1}$ in $B[1,i]$, and (iii)  $select_b(B,j)$ returns the position in $B$ of the $j^{th}$ occurrence of bit $b\in{0,1}$. There exist representations using $n+o(n)$ bits that answer all those queries in constant time \cite{Clark:1996}. When the bitmap has $m \ll n$ 1s, it is possible to use compressed representations that use $m\log(n/m)+O(m)$ bits \cite{Eli74,Fan71}. This representation still performs $select_1$ queries in constant time, whereas $access$ and $rank$ require time $O(\log(n/m))$ \cite{Okanohara:2007:PER:2791188.2791194}.

\medskip
\noindent
\textbf{Partial sums.} Given values $0 < x_1 < x_2 < \ldots < x_m \le n$, we can define the differences $d_i =
x_i-x_{i-1}$ and $d_1=x_1$, so that $x_i = \sum_{j=1}^i d_i$. An Elias-Fano representation of the partial sums
is a bitmap $B[1..n]$ with all $B[x_i] = 1$ and all the rest zero, or which is the same, the concatenation of the $d_i$ values written in unary. Therefore, we can retrieve $x_i =
select_1(B,i)$ in constant time, and the space of the representation is $\log(n/m)+O(m)$ bits, close to a differential representation of the $d_i$ values.

\section{Related work}

\noindent
\textbf{Reducing the size of trajectories.}
A lossy way to reduce size is to generate a new trajectory that approximates the original one, by keeping the most representative points. The best known method of this type is the Douglas-Peucker algorithm \cite{Douglas:1973:ARN}. Other strategies record speed and direction, discarding points that can be reasonably predicted with this data \cite{TrajcevskiCSWV06}.
A lossless way to reduce space is to use differential encodings of the consecutive values $x$, $y$, and time \cite{TrajStore,Trajic,Wang:2014}. 

\medskip
\noindent
\textbf{Spatio-temporal indexes.}
Spatio-temporal  indexes can be classified into three types. 
  The first is a classic multidimensional spatial index, usually the R-tree, augmented  with a temporal dimension. For example, the 3DR-tree \cite{Vazirgiannis1998} uses three-dimensional Minimum Bounding Boxes (MBBs), where the third dimension is the time, to index segments of trajectories.
A second approach is the multiversion R-trees, which creates an R-tree for each timestamp and a B-tree to select the relevant R-trees. The best known index of this family is the MV3R-tree \cite{PapadiasT01}.
The third type of index partitions the space statically, and then a temporal index is built for each of the spatial partitions \cite{ChakkaEP03}.


\subsection{GraCT}\label{Grct}

The closest predecessor of our work,
GraCT \cite{BrisaboaGNP16}, assumes regular timestamps and stores trajectories using two components. At regular time instants, it represents the position of all the objects in a structure called \textit{snapshot}. The positions of objects between snapshots are represented in a structure called \textit{log}.

Let us denote $Sp_{k}$ the snapshot representing the position of all the objects at timestamp $k$. Between two consecutive snapshots $Sp_{k}$ and $Sp_{k+d}$, there is a log for each object, which is denoted ${\cal L}_{k,k+d}(id)$, being $id$ the identifier of the object. 
 The log stores the differences of positions compressed with {\em RePair}  \cite{larsson2000off}, a grammar-based compressor. In order to speed up the queries over the resulting sequence, the nonterminals are enriched with additional information, mainly the MBB of the trajectory segment encoded by the nonterminal. 

 Each snapshot is a binary matrix where a cell set to 1 indicates that one or more objects are placed in that position of the space. To store such a (generally sparse) matrix, it uses a $k^2$-tree \cite{ktree}. The $k^2$-tree is a space- and time- efficient version of a region quadtree \cite{Sam2006}, and is used to filter the objects that may be relevant for a time-instant or time-interval query.

%
%
%

\subsection{ScdcCT}
ScdcCT was implemented as a classical compressed baseline to compare against GraCT \cite{BrisaboaGNP16}. It uses the same components, snapshots and logs, but the logs are compressed with differences and not with grammars. The differences are compressed using $(s,c)$-Dense Codes \cite{RodrguezBrisaboa07}, a fast-to-decode variable-length code that has low redundancy over the zero-order empirical entropy of the sequence. This exploits the fact that short movements to contiguous cells are more frequent than movements to distant cells. 

\section{ContaCT}

ContaCT uses snapshots and logs, just like GraCT.
The main differences are in the log. As explained, in GraCT the log stores the differences of the consecutive positions. In order to know the position of an object at a given timestamp $i$,  we access the closest previous snapshot and add up the differences until reaching the desired timestamp. GraCT speeds up this traversal by storing the total differences represented by nonterminals, so that they can be traversed in constant time. This makes GraCT faster than a differential representation that needs to add up all the individual differences, but still it has to traverse a number of symbols that grows proportionally to the distance $d$ between consecutive snapshots. ContaCT completely avoids that sequential traversal of the log.

\subsection{The log}


 ContaCT represents each ${\cal L}_{k,k+d}(id)$ with components $time(id)$, $\Delta_X(id)$, $\Delta_Y(id)$.
 
Time $(id)$ tells the timestamps for which object $id$ has $(x,y)$ coordinates. It stores the $first$ and $last$ positions with data in ${\cal L}_{k,k+d}(id)$, and a bitmap $T(id)$ of $last-first+1$ bits indicating with a 0 that there is data at that time instant.
 
$\Delta_{X}(id)$ stores the differences of the $x$ coordinate using three bitmaps: $X(id)_t$ indicates, for each position having a 0 in $T(id)$, whether the difference is positive or negative;
and $X(id)_{p}$ and $X(id)_n$ store the positive and negative differences, respectively, using Elias-Fano. $\Delta_{Y}(id)$ is analogous.

Given the log ${\cal L}_{k,k+d}(id)$ and a local timestamp $i\in [1,d-1]$, 
we compute the $x$ coordinate of the object $id$ at that timestamp as follows (analogous for $y$):

\vspace{-2mm}
\begin{enumerate}
\item $dis=rank_1(T(id),i-first+1)$ returns the number of timestamps for which we have no data (the object was missing) until position $i$, counting from the first timestamp with data.
\item $pos=rank_1(X(id)_t,i-dis-first+1)$ and $neg=i-dis-first-pos+1$, are the number of positive and negative differences until position $i$, respectively.
\item $select_1(X(id)_p,pos)-pos-(select_1(X(id)_n,neg)-neg)$ returns the $x$ coordinate at timestamp $i$.
\end{enumerate} 

\vspace{-2mm}

We use the sparse bitmap representation for $X(id)_{p}$ and  $X(id)_{n}$, and the plain version for $X(id)_t$ and $T(id)$. The size of the complete structure is  $n \log N/n + O(d)$ bits, where $N$ is the sum of the differences in $x$, and $n \le d$ is the number of positions where the object has coordinate information.

%
 
\begin{figure}[t]
\begin{center}

\includegraphics[scale=0.45]{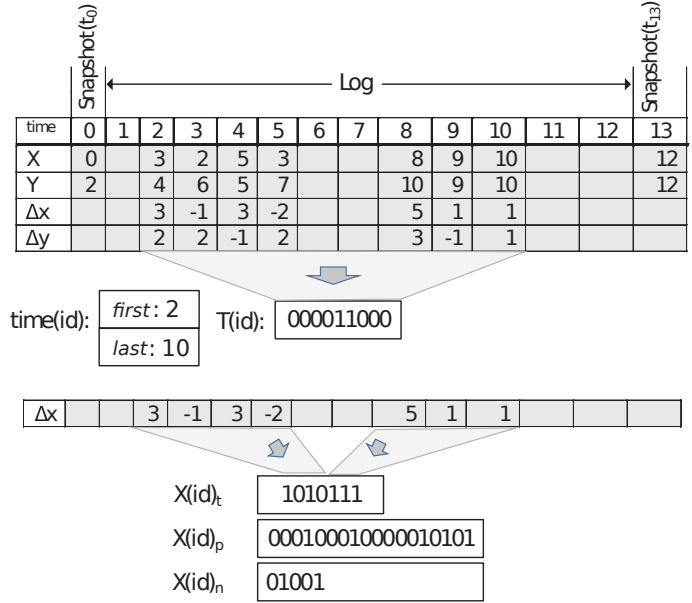}
\end{center}
\vspace*{-7mm}
\caption{The log of ContaCT for a given object \textit{id}.}\label{estructura}
\end{figure}

\medskip
\noindent
\textbf{Example.}
The top of Figure \ref{estructura} shows the coordinates of a trajectory. There is no data about the position of the object at timestamps 1, 6, 7, 11, and 12.
Timestamps $0$ and $13$ are represented with snapshots. Arrays $X$ and $Y$ contain the absolute coordinates of the trajectory, and $\Delta X$ and $\Delta Y$ the corresponding differences (the arrays are not stored in this form,  they are included for clarity). 

Below those arrays, we have the data structure $time(id)$: {\em First} and {\em last} store the first and last timestamps of ${\cal L}_{0,13}(id)$ that have data, and bitmap $T(id)$ has a bit for each timestamp in between. A bit 1 means no data for its timestamp.

The bottom of the figure shows the three bitmaps that represent $\Delta X(id)$. $X(id)_t$ has a bit for each bit set to 0 in $T(id)$, that is, for each position of $\Delta X(id)$ with a value. Each bit of $X(id)_t$ indicates whether the corresponding difference is positive or negative. For each bit of $X(id)_t$ set to 1, $X(id)_p$ stores that value in unary. $X(id)_n$ stores, in the same way, the negative differences.

Let us extract the $x$ coordinate at timestamp $9$. First, we obtain the number of disappearances until timestamp $9$: $dis=rank_1(T(id),i-first+1)=2$. Next, we obtain the number of positive and negative differences until timestamp $9$: $pos=rank_1(X(id)_t,i-dis-first+1)=4$ and $neg=i-dis-first-pos+1=2$. Finally, the $x$ coordinate is $select_1(X(id)_p,pos)-pos-(select_1(X_n,neg)-neg)= select_1(X(id)_p,4)-4-(select_1(X(id)_n,2)-2)$=$16-4-(5-2)=12-3=9$.
\qed 

\subsection{Indexing the logs}

Our representation yields constant-time extraction of whole trajectories and direct access to any point. To solve time-slice and time-interval queries, we may just compute the position or consecutive positions of the object and see if they fall within the query area. Although we can rapidly know the position of an object in a given timestamp, if we have to inspect all the timestamps of a given queried interval, we may spend much time obtaining positions that are outside the region of interest. In order to accelerate these queries over the logs, ContaCT stores an index for each ${\cal L}_{k,k+d}(id)$.

The index is a perfect binary tree that indexes the timestamps of the interval $[k+1,k+d-1]$ containing data (i.e., after being mapped with $T(id)$). Let $C$ indicate the number of timestamps covered by a leaf. Internal nodes cover the ranges covered by all the leaves in their subtree. Each node stores the MBR of the positions of the object during their covered interval of timestamps. 

To check the positions of the object in the interval $[b,e]$, where $1 \leq b\leq e< d$, we first compute $b'=rank_0(T(id),b-first)$ and $e'=rank_0(T(id),e-first)$, and then check the timestamps of the tree in the range $[b',e']$. The way to use this tree is described in the next subsection.


\begin{figure}[t]
\begin{center}

\includegraphics[scale=0.6]{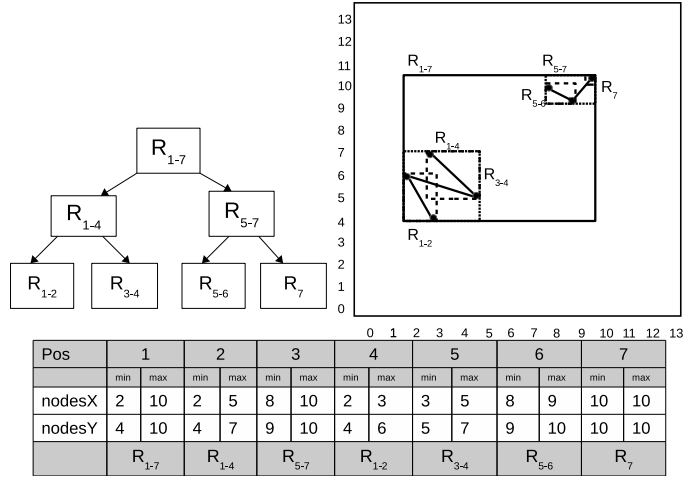}
\end{center}
\vspace*{-5mm}
\caption{The index of a log.}\label{index}
\end{figure}

\medskip
\noindent
\textbf{Example.}
Figure \ref{index} shows the index for the trajectory of Figure \ref{estructura}. $C$ is 2, so the leaves cover at most 2 timestamps.  In ${\cal L}_{0,13}(id)$, there are 7 time instants with values, at timestamps 2, 3, 4, 5, 8, 9, and 10. Therefore the leftmost leaf of the tree covers the positions at timestamps 2 and 3, the next leaf covers the timestamps 4 and 5, and so on.

 The root contains the MBR that encloses all the positions in the interval covered by ${\cal L}_{0,13}(id)$. Since there are 7 timestamps with values, we labeled it $R_{1-7}$. At the top right, that MBR is drawn as a rectangle with a solid line. The left child, $R_{1-4}$, covers the positions of the first 4 timestamps. The right child, $R_{5-7}$, covers the timestamps from the $5^{th}$ to the $7^{th}$, and so on. The second-level MBRs are shown at the top right  as rectangles with densely dotted lines, whereas the third level MBRs are drawn with scattered dotted lines. 
\qed \medskip

Observe that each log stores the movements of one object between two snapshots, therefore there will be a considerable number of trees. To save space, we store the perfect trees in heap order, avoiding pointers. Each tree is then stored as two arrays, {\em nodesX} and {\em nodesY}, storing the extremes of the MBRs in each dimension. The children of a node at position $p$ are at $2p$ and $2p+1$.

 
Further, the arrays {\em nodesX} and {\em nodesY} are compressed by storing the values of the nodes below the root as differences with respect to their parent. For example, the values at  position 2 (corresponding to $R_{1-4}$) of {\em nodeX} are stored as the values of the parent (2,10) minus the values at position 2 (2,5), that is, (0,5). As a result, the numbers are smaller, and we use $\lfloor \log m \rfloor+1$ bits for each number, being $m$ the largest difference (the root MBRs are stored separately).
 
 \subsection{Queries}

To answer a time-slice or a time-interval query, we use the closest previous snapshot to filter the objects that cannot possibly make it to the query region within the given time frame, by exploiting the maximum speed at which objects can move. Let $r = [x_1,x_2] \times [y_1,y_2]$ be a rectangular region in the two-dimensional space, and $b<e$ be two timestamps. Let $s$ be the maximum speed, in our dataset, of any object. We denote $ER(r,q)$, the \textit{expanded region} of $r$ at timestamp $q$, the area that contains the points that must be considered from the preceding snapshot. If the timestamp of the preceding snapshot is $k$, then $ER(r,q) = [x_1-s\cdot(q-k),x_2+s\cdot(q-k)] \times [y_1-s\cdot(q-k),y_2+s\cdot(q-k)]$.

 \medskip
 \noindent
 \textbf{Time-slice.}
 A time-slice query specifies a region $r$ and a timestamp $q$. Assume $q$ is between snapshots $Sp_k$ and $Sp_{k+d}$. We perform a range query on $Sp_k$ to retrieve all the objects $id$ in $ER(r,q)$. If $q=k$, we simply return all those objects $id$. Otherwise, we access the log $\mathcal{L}_{k,k+d}$ of each such object $id$ to find, in $O(1)$ time, its position at (local) time $q-k$, and report $id$ if the position is within $r$.

 \medskip
 \noindent
 \textbf{Time-interval.}
A time-interval query specifies a region $r$ and an interval  $[b,e]$. It can be solved as a sequence $e'-b'+1$ time-slice queries (where $b'$ and $e'$ are described previously), but we exploit the tree of MBRs to speed up the query.

Each object that is within $ER(r,q)$ must be tracked along the timestamps $b$ to $e$, to determine if it has a position inside $r$. We compute $b'$ and $e'$ as described previously and use the MBR tree to quickly filter out the elements that do not qualify. We start at the tree root, and check if (1) the timestamps of the node intersect $[b',e']$ and (2) the root MBR intersects $r$. If not, we abandon the search at that node. Otherwise, we recursively enter its left and right children. When we reach a leaf, we extract all the positions one by one, looking for the first that falls within $r$. We develop specialized procedures to extract the next point faster than a random access in our Elias-Fano representation.

We further prune the search by continuously considering the maximum speed of the objects. Assume $[b',e']$ is within the right child of a node since the left one covers only $[b_1',e_1']$. If the minimum distance between the MBR of the left and $r$, along any coordinate, is $p > s \cdot(b-e_1)$, then there is no need to examine the right child. Here $e_1$ is the original timestamp corresponding to $e_1'$, which is obtained with $select_0(T(id),e_1')+first-1$. The same argument holds symmetrically with the left child. Finally, as we traverse the positions in a leaf, we verify this condition continuously to preempt the scan as soon as possible (we use a special ``select-next'' method on $T(id)$ to speed up consecutive $select$ queries).

\medskip
\noindent
\textbf{ Example.}
Let us run the time-interval query for the area $r=[4,5] \times [4,10]$ and (mapped) time interval $[b',e'] = [2,4]$ in the log of Figure~\ref{index}. We start at the root, which covers the time range $[1,7]$ and has MBR $[2,10]\times[4,10]$. Since both intersect the query, we continue. Since the tree is perfect, we know that the left subtree covers the timestamps $[1,4]$ and the right one covers $[5,7]$. Since the right child does not intersect the query time interval, we only descend by the left one, $R_{1,4}$. Its MBR is $[2,5] \times [4,7]$, which intersects $r$, so we continue. Its left child, $R_{1,2}$, covers the time interval $[1,2]$, which intersects $[b',e']$, so we enter it. However, its MBR is $[2,3] \times [4,6]$, which does not intersect $r$ and thus we abandon it. The right child of $R_{1,4}$, $R_{3,4}$, also intersects the time interval of the query. Its MBR is $[3,5] \times [5,7]$, which intersects $r$. Finally, since $R_{3,4}$ is a leaf, we access the $3^{rd}$ and $4^{th}$ positions in the log, finding that the object was in $r$ at time instant $4$.
\qed 
%

\section{Experimental Results}





ContaCT was coded in C++  and  uses several data strucures of the SDSL library\cite{gbmp2014sea}. As baselines, we include GraCT and ScdcCT \cite{BrisaboaGNP16}, also C++ programs, and the MVR-tree from the spatialindex library (\texttt{libspatialindex.github.io}). We used a real dataset storing the movements of $3{,}654$ ships on a grid of size $2{,}723 \times 367{,}775$ and $44{,}642$ time instants, whose plain representation requires 395.07 MB; we measure our compression ratios against that size. Appendix \ref{appendix:dataset} gives more details on the dataset.

The experiments ran on an Intel\textsuperscript{\textregistered} Core\textsuperscript{TM} i7-3820 CPU @ 3.60GHz (4 cores) with 10MB of cache and 64GB of RAM, over Ubuntu 12.04.5 LTS with kernel 3.2.0-115 (64 bits), using gcc 4.6.4 with \texttt{-O9}.
We tested six types of queries:
\begin{itemize}
\item {\em Object} searches for the position of a specific object at a given timestamp.
\item {\em Trajectory} returns the positions of an object between two timestamps.
\item {\em Slice S} and {\em Slice L} are time-slice queries for small regions ($272 \times 367$\ cells) and large regions ($2723 \times 3677$\ cells), respectively.
\item {\em Interval S} are time-interval queries specifying a small region on small intervall (36 timestamps), and {\em Interval L} are time-interval queries specifying large regions on large intervals (90 timestamps).
\end{itemize}   

 We measure elapsed times.
Each data point averages 20{,}000 Object queries, 10{,}000 Trajectory queries, or 1{,}000 of Slice/Interval queries.

\begin{figure}[t]
	\centering     
    \subfigure[\texttt{Size}]
{\label{fig:size}\includegraphics[width=0.49\textwidth]{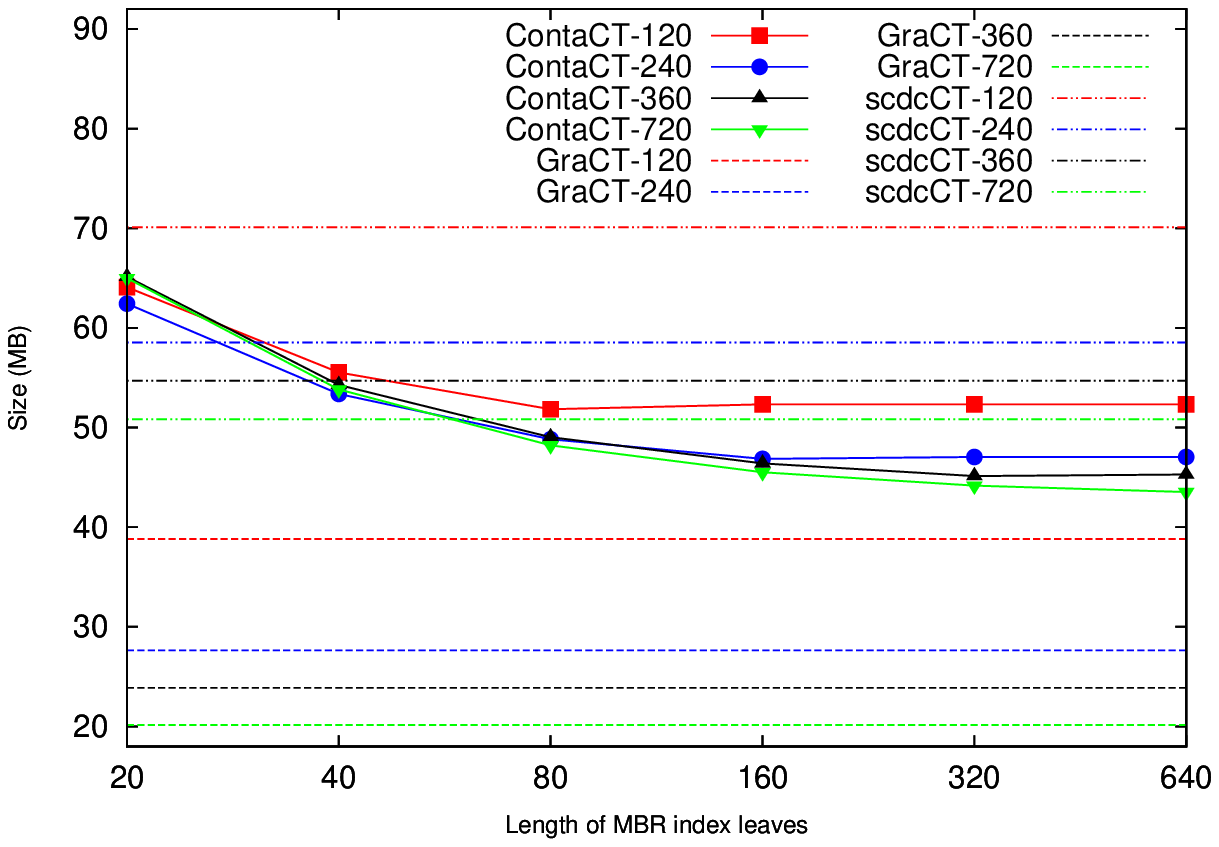}}
\subfigure[\texttt{Object and trajectory}]
{\label{fig:time-ot-trajectory}\includegraphics[width=0.49\textwidth]{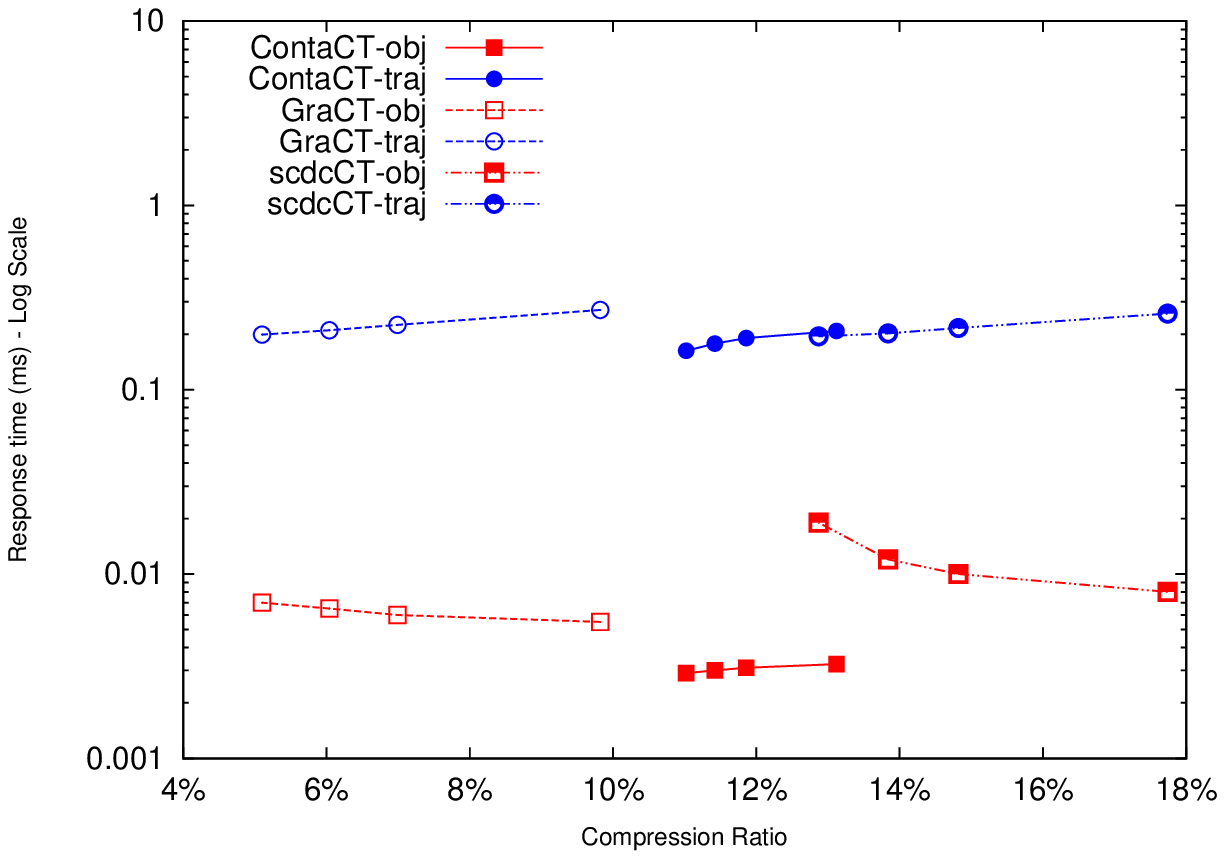}}
\subfigure[\texttt{Slice S and Slice L}]
{\label{fig:time-slice}\includegraphics[width=0.49\textwidth]{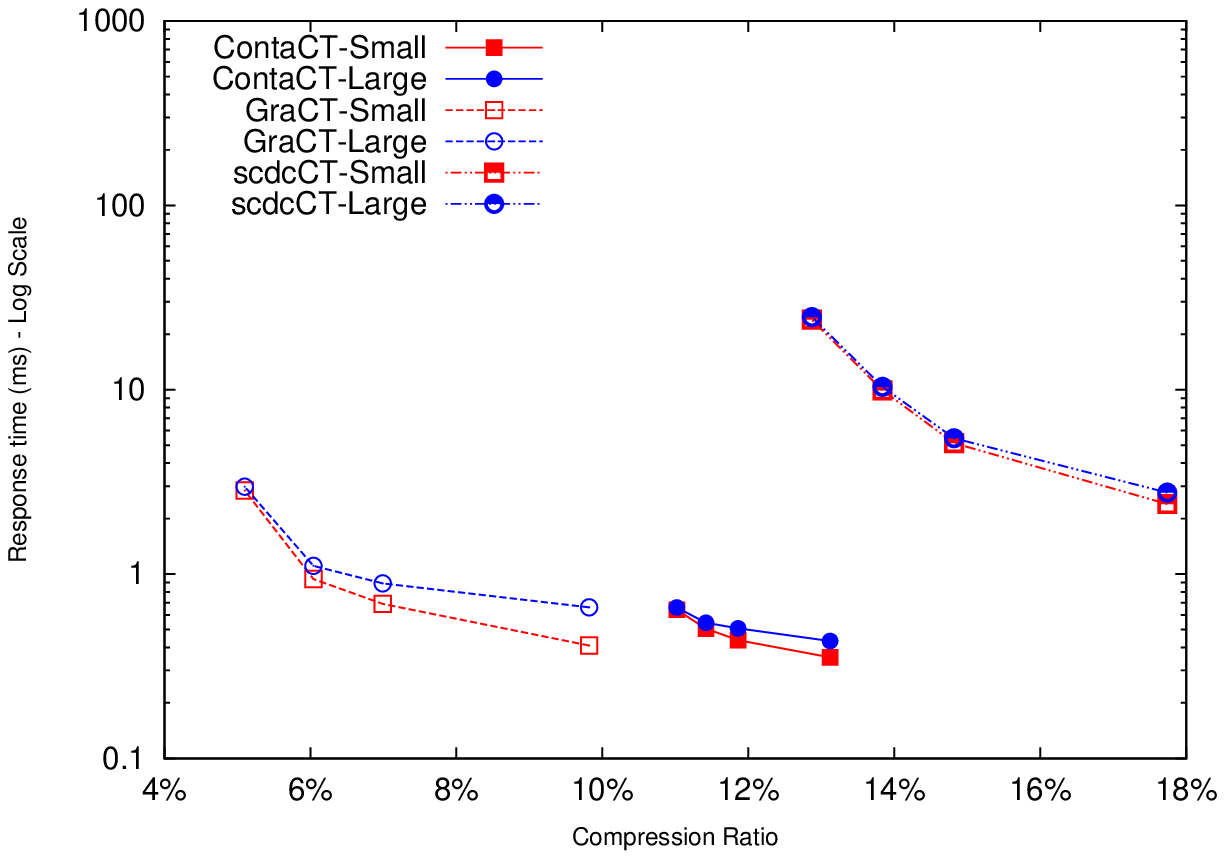}}
\subfigure[\texttt{Interval S and Interval L}]
{\label{fig:time-interval}\includegraphics[width=0.49\textwidth]{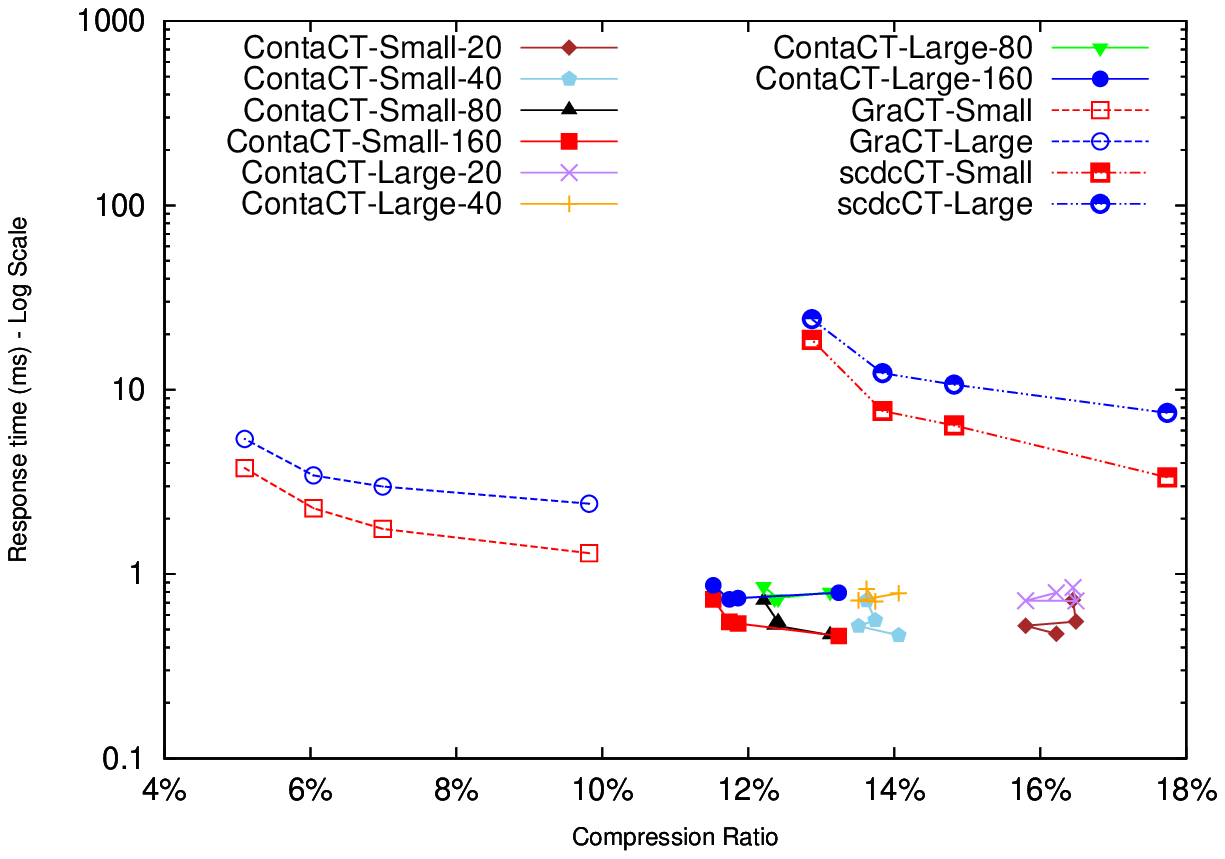}}

\caption{Compressed representations comparison; beware of logarithmic $y$ axes.}
\label{fig:experiment2}
\end{figure}

\medskip
\noindent
\textbf{Compressed representations}. We built ContaCT, ScdcCT and GraCT with different snapshot distances, namely every 120, 240, 360, and 720 timestamps. ContaCT was also built with different values of $C$ (the number of timestamps covered by the leaves of the MBR trees), specifically  20, 40, 80, 160, 320, and 640. We used Elias-Fano on the bitmaps $T(id)$, which were sparse, but turned to plain bitmaps to represent $X(id)$ and $Y(id)$, as they were not sufficiently sparse after mapping from $T(id)$.

Figure \ref{fig:size} shows the size with the different settings. All the structure sizes decrease as the distances between snapshots increase, and ContaCT also decreases as $C$ increases. Thanks to its grammar-compression, the densest snapshot sampling of GraCT still uses 11\% less space than the sparsest sampling of ContaCT. In turn, ContaCT is smaller than the other differentially compressed representation, ScdcCT, for example by 14\% in their sparsest configurations.
 

Figure \ref{fig:time-ot-trajectory} shows the average answer times for Object and Trajectory queries. ContaCT is especially fast on Object queries, thanks to its constant-time extraction mechanism. This makes it mostly independent of the snapshot sampling, and twice as fast as GraCT and three times faster than ScdcCT, even with their fastest configurations. GraCT is faster than ScdcCT, because it can traverse nonterminals of the grammar in constant time. For Trajectory queries, ContaCT is still faster by 20\%. The difference decreases because sequential access to trajectories is not comparatively that slow with the other methods. The reason why some curves actually improve with a sparser snapshot sampling is that some extra work is needed when the query goes through various snapshots.

Figure \ref{fig:time-slice} shows time-slice queries. The snapshot sampling is now crucial, since it affects the number of candidates that must be considered from the preceding snapshot (the computation of $ER(r,q)$). Since ContaCT can access the desired time instant in constant time, it is considerably faster than the others for a given snapshot sampling. However, GraCT  matches ContaCT (and outperforms it for more selective queries) for a similar space usage, because GraCT can use a denser sampling thanks to its better compression of the log. ContaCT, on the other hand, outperforms ScdcCT by far.

Figure \ref{fig:time-interval} shows time-interval queries, with various values of $C$ for ContaCT.
Even with the nearly smallest-space configuration (snapshot interval 360, $C=160$), ContaCT outpeforms the largest GraCT configuration by a factor of 2, thanks to the MBR trees that index the logs. Using smaller $C$ values does not significantly improve the time, on the other hand, thanks to our optimized leaf traversal procedure.  Once again, the baseline ScdcCT is much slower.


\medskip
\noindent
\textbf{Comparison with a spatio-temporal index.} We compare ContaCT with MVR-tree, a classic spatio-temporal index. We configured MVR-tree to run in main memory. To avoid  space problems, we had to build the MVR-tree over a quarter of the input dataset. The size of the MVR-tree on this reduced input was 15.41 GB (including the data), while the maximum-space configuration of ContaCT uses 11.61 MB, three orders of magnitude less.

\begin{figure}[t]
\centering     
    \subfigure[\texttt{Slice and Interval queries}]
{\label{fig:mvrtree1}\includegraphics[width=0.49\textwidth]{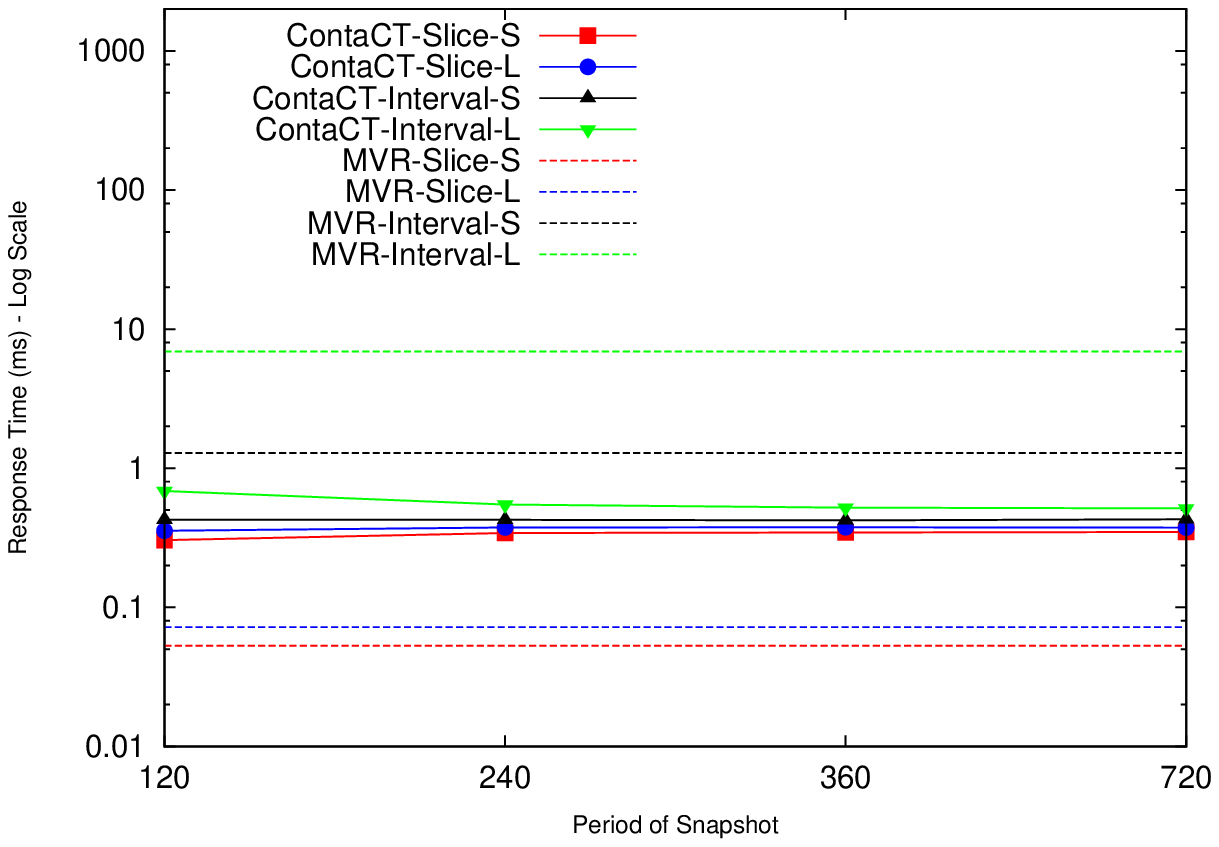}}
\subfigure[\texttt{Growing time-interval queries}]
{\label{fig:mvrtree2}\includegraphics[width=0.49\textwidth]{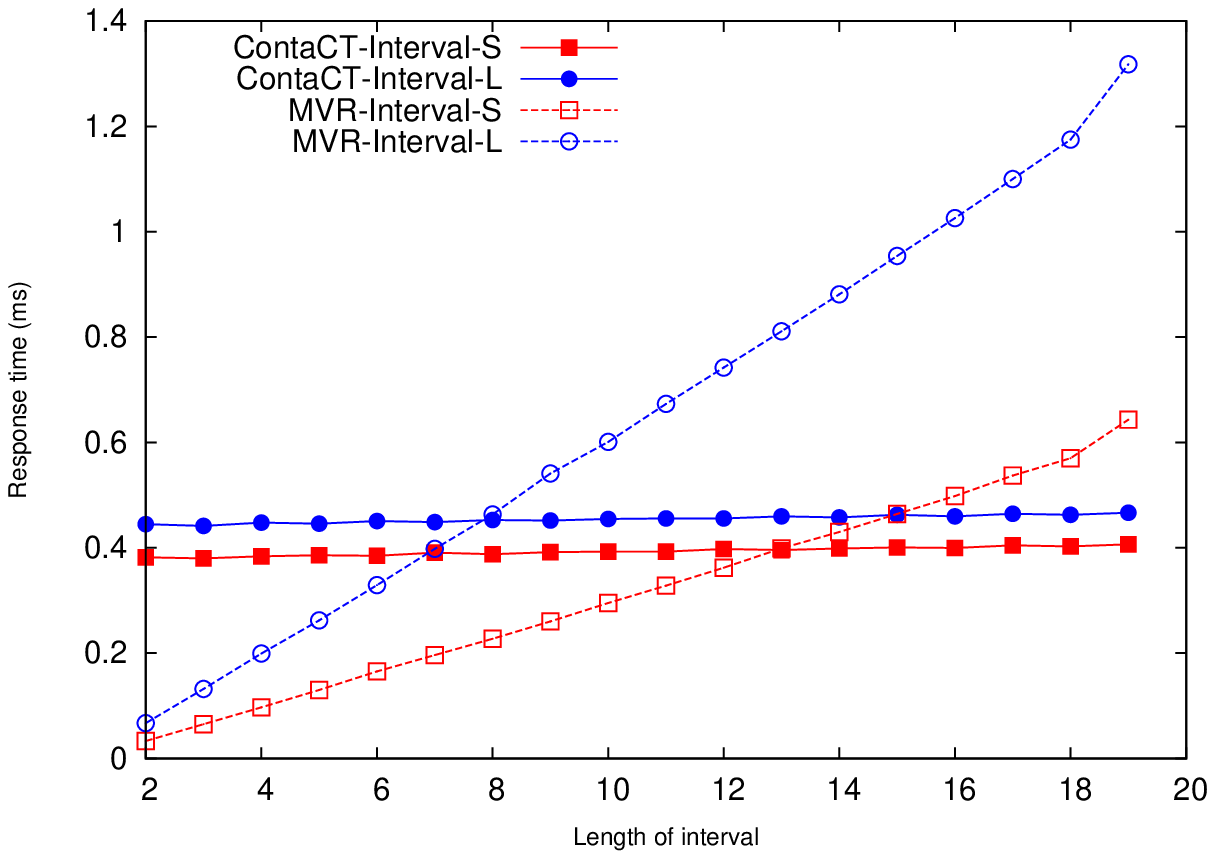}}
\caption{Comparison with spatio-temporal index MVR-tree.}
\label{fig:mvrtree}
\end{figure}

The MVR-tree can only solve time-slice and time-interval queries.  We built ContaCT with different snapshot samplings and $C=80$.  Figure \ref{fig:mvrtree1} shows that our structure is faster on time time-interval queries, but slower on our time-slice queries.
 Figure \ref{fig:mvrtree2} studies the turning point, by increasing the time span of time-interval queries, using the smallest-space configuration of ContaCT (snapshot period of 720). Note that MVR-tree times increase linearly whereas ContaCT stays essentially constant. ContaCT outperforms MVR-tree on interval lengths over 8 on large-region queries and over 14 in small-region ones. 


\section{Conclusions}

We have presented ContaCT, a structure to index trajectories of sets of moving objects in compressed form. ContaCT can efficiently retrieve points or segments of individual trajectories, and answer spatio-temporal range queries on the set of objects. ContaCT combines sampled two-dimensional snapshots compressed with $k^2$-trees, with logs differentially compressed and represented with Elias-Fano, which gives constant-time access to trajectory points. It also includes a hierarchical MBR mechanism that, combined with a pruning done on the snapshots, efficiently answers spatio-temporal queries.
 
Our experiments show that ContaCT compresses the data by a factor of almost 10 and outperforms by far, in space and time, a baseline alternative based on compressing small consecutive differences. ContaCT is also more than 1{,}000 times smaller than a classical spatio-temporal index, while being faster on all but very time-narrow queries. Compared with GraCT, the smallest existing representation based on grammar-compressing the trajectories, ContaCT uses more space. However, when both indexes are set to use the same amount of space, ContaCT generally makes better use of it, outperforming GraCT in most queries, by a factor of up to 3.

Future work involves extending ContaCT to more sophisticated queries, such as nearest-neighbor spatio-temporal queries.

\bibliographystyle{splncs03}
\bibliography{bibliografia}

\begin{thebibliography}{10}
\providecommand{\url}[1]{\texttt{#1}}
\providecommand{\urlprefix}{URL }

\bibitem{RodrguezBrisaboa07}
Brisaboa, N.R., Fari\~na, A., Navarro, G., Paramá, J.R.: Lightweight natural
  language text compression. Information Retrieval  10(1),  1--33 (2007)

\bibitem{ktree}
Brisaboa, N.R., Ladra, S., Navarro, G.: Compact representation of web graphs
  with extended functionality. Information Systems  39(1),  152--174 (2014)

\bibitem{BrisaboaGNP16}
Brisaboa, N.R., G{\'{o}}mez{-}Brand{\'{o}}n, A., Navarro, G., Param{\'{a}},
  J.R.: Gract: {A} grammar based compressed representation of trajectories. In:
  SPIRE. pp. 218--230 (2016)

\bibitem{ChakkaEP03}
Chakka, V.P., Everspaugh, A., Patel, J.M.: Indexing large trajectory data sets
  with {SETI}. In: CIDR (2003)

\bibitem{Clark:1996}
Clark, D.: Compact Pat Trees. Ph.D. thesis, Univ. Waterloo (1996)

\bibitem{TrajStore}
Cudre-Mauroux, P., Wu, E., Madden, S.: Trajstore: An adaptive storage system
  for very large trajectory data sets. In: ICDE. pp. 109--120 (2010)

\bibitem{Douglas:1973:ARN}
Douglas, D.H., Peuker, T.K.: Algorithms for the reduction of the number of
  points required to represent a line or its caricature. The Canadian
  Cartographer  10(2),  112--122 (1973)

\bibitem{Eli74}
Elias, P.: Efficient storage and retrieval by content and address of static
  files. Journal of the ACM  21,  246--260 (1974)

\bibitem{Fan71}
Fano, R.: On the number of bits required to implement an associative memory.
  Memo 61, Computer Structures Group, Project MAC, Massachusetts (1971)

\bibitem{gbmp2014sea}
Gog, S., Beller, T., Moffat, A., Petri, M.: From theory to practice: Plug and
  play with succinct data structures. In: SEA. pp. 326--337 (2014)

\bibitem{larsson2000off}
Larsson, N.J., Moffat, A.: Off-line dictionary-based compression. Proceedings
  of the IEEE  88(11),  1722--1732 (2000)

\bibitem{Trajic}
Nibali, A., He, Z.: Trajic: An effective compression system for trajectory
  data. IEEE Transactions on Knowledge and Data Engineering  27(11),
  3138--3151 (2015)

\bibitem{Okanohara:2007:PER:2791188.2791194}
Okanohara, D., Sadakane, K.: Practical entropy-compressed rank/select
  dictionary. In: ALENEX. pp. 60--70 (2007)

\bibitem{PfoserJT00}
Pfoser, D., Jensen, C.S., Theodoridis, Y.: Novel approaches to the indexing of
  moving object trajectories. In: VLDB. pp. 395--406 (2000)

\bibitem{Sam2006}
Samet, H.: {Foundations of Multimensional and Metric Data Structures}. Morgan
  Kaufmann (2006)

\bibitem{PapadiasT01}
Tao, Y., Papadias, D.: {MV3R}-tree: {A} spatio-temporal access method for
  timestamp and interval queries. In: VLDB. pp. 431--440 (2001)

\bibitem{TrajcevskiCSWV06}
Trajcevski, G., Cao, H., Scheuermann, P., Wolfson, O., Vaccaro, D.: On-line
  data reduction and the quality of history in moving objects databases. In:
  MobiDE. pp. 19--26 (2006)

\bibitem{Vazirgiannis1998}
Vazirgiannis, M., Theodoridis, Y., Sellis, T.K.: Spatio-temporal composition
  and indexing for large multimedia applications. ACM Multimedia Systems
  Journal  6(4),  284--298 (1998)

\bibitem{Wang:2014}
Wang, H., Zheng, K., Xu, J., Zheng, B., Zhou, X., Sadiq, S.: Sharkdb: An
  in-memory column-oriented trajectory storage. In: CIKM. pp. 1409--1418 (2014)

\bibitem{zheng11}
Zheng, Y., Zhou, X. (eds.): Computing with Spatial Trajectories. Springer
  (2011)

\end{thebibliography}

\appendix

\section{Dataset details}
\label{appendix:dataset}

The dataset used in our experimental evaluation corresponds to a real dataset storing the movements of 3,654 boats sailing in the UTM Zone 10 during one month of 2014. It was obtained from MarineCadastre.\footnote{\texttt{http://marinecadastre.gov/ais/}} Every position emitted by a ship is discretized into a matrix where the cell size is $50 \times 50$ meters. With this data normalization, we obtain a matrix with 1,001,451,325 cells, 2,723 in the $x$-axis and 367,775 in the $y$-axis. As our structure needs the position of the objects at regular timestamps, we preprocessed the signals every minute, sampling the time into 44,642 minutes in one month. 

To filter out some obvious GPS errors, we set the maximum speed of our dataset to 55 cells per minute (over 234 km/h) and deleted every movement faster than this speed. In addition, we observe that most of the boats sent their positions frequently when they were moving, but not when they were stopped or moving slowly. This produced logs of boats with many small periods without signals (absence period). Taking into account that an object cannot move too far away during a small interval of time, we interpolated the signals when the absence period was smaller than 15 minutes, filling the periods of absence with these interpolated positions.

With these settings the original dataset occupies 974.43 MB in a plain text file with four columns: \textit{object identifier}, \textit{time instant}, \textit{coordinate x} and \textit{coordinate y}. Every value of these columns are stored as a string. However, to obtain a more precise compression measure, we represent this information in a binary file using two bytes to represent object identifiers (max value 3,653), two bytes for the instant column (max value 44,641), two bytes for the x-axis (max value 2,723) and three bytes for the y-axis (max value 367,775). Therefore, the binary representation of our dataset occupies 395.07MB.

\end{document}